\documentclass[preprint]{revtex4-1}
\usepackage{titlesec}
\usepackage{lipsum}

\titlespacing*{\section}
{0pt}{0ex}{1ex}
\titlespacing*{\subsection}
{0pt}{0ex}{-1ex}
\titlespacing*{\subsubsection}
{0pt}{0ex}{-2ex}

\usepackage{color}
\usepackage{graphicx}
\usepackage{natbib}
\usepackage{epsfig}
\usepackage{setspace}
\usepackage{amsmath}
\usepackage{amssymb}
\usepackage{verbatim}
\usepackage{bibentry}

\RequirePackage[left=0.6in,right=1.in,top=1in,bottom=1in]{geometry}  
\usepackage{array}
\newcolumntype{P}[1]{>{\raggedright\let\newline\\\arraybackslash\hspace{0pt}}m{#1}}
\newcolumntype{L}[1]{>{\raggedright\let\newline\\\arraybackslash\hspace{0pt}}m{#1}}
\newcolumntype{C}[1]{>{\centering\let\newline\\\arraybackslash\hspace{0pt}}m{#1}}
\newcolumntype{R}[1]{>{\raggedleft\let\newline\\\arraybackslash\hspace{0pt}}m{#1}}
\begin{document}

\begin{titlepage}
\title{Schrieffer-Wolff transformation of Anderson Models}

\author{Rukhsan Ul Haq}
\affiliation{Theoretical Sciences Unit, Jawaharlal Nehru Center for Advanced Scientific Research}

\author{Sachin Satish Bharadwaj}
\email{satish@jncasr.ac.in}
\affiliation{Theoretical Sciences Unit, Jawaharlal Nehru Center for Advanced Scientific Research\\Department of Mechanical Engineering, RV College of Engineering\\Theoretical Physics and Mathematics, Centre for Fundamental Research and Creative Education, Bangalore India}

\begin{abstract}
{ Schrieffer-Wolff transformation is one of the very important transformations in the study of quantum many body physics. It is used to arrive at the low energy effective hamiltonian of  Quantum many-body hamiltonians, which are not generally  analytically tractable. In this paper we give a pedagogical review of this transformation for Anderson Impurity model(SIAM) and its lattice generlaization called, the Periodic Anderson Model(PAM).}
\end{abstract}
\maketitle
\end{titlepage}

\section{INTRODUCTION}
In the study of quantum many-body physics, the hamiltonians are in general, very difficult to be treated analytically. Thus, one tries to understand the low energy spectrum of the underlying physics by computing, effective hamiltonians of the quantum many-body system. One of the prominent examples of this approach is the Kondo model, which is essentially the low energy effective hamiltonian of the Anderson Impurity model(SIAM). The Kondo model which captures the low energy physics of Anderson Impurity model, is quite helpful in understanding the Kondo physics itself which is a many-body pheneomena. Schrieffer-Wolff(SW) transformation was historically used to get the Kondo model from SIAM. Schrieffer-Wolff transformation is a method to arrive at effective hamiltonians and has been extensively used both in quantum mechanics and quantum many-body physics. SW transformation can also be used to diagonalize a given hamiltonain, as it forms a unitary transformation, yet it may not lead to a full-fledged diagonalization; rather, in many cases, it just reduces the hamiltonian to its corresponding band diagonal form. Yet another use of SW transformation is, to view the formalism as a degenerate perturbation theory. Historically SW transformation was used to solve many important problems and has been given different names such as Frohlich transformation in the electron-phonon problem, Foldy-Wouthuysen transformation in relativistic Quantum mechanics and k.p perturbation theory in semiconductor physics. SW transformation was generalized by Wegner, as the Flow equation method, which aids in circumventing the divergences from vanishing denominators, and also has been  quite a successful method in various other problems governing non-equilibrium physics as well.

The Schrieffer-Wolff transformation, as mentioned earlier, is basically a unitary transformation. It chooses a unitary operator to obtain the transformed hamiltonian.
\begin{align}
H'&=U^{\dag}HU \\
H'&=e^{-S}He^{S} \\
&= H_{0}+\frac{1}{2}\left[S,H_{v}\right]+\frac{1}{3}\left[S,\left[S,H_{v}\right]\right]+...
\end{align}
where, S is called generator of the SW transformation. Usually $H$ can be written as $H=H_{0}+H_{v}$ , where $H_{0}$ is the diagonal part and $H_{v}$ is the off-diagonal part of the hamiltonian. The generator $S$ is chosen in such a way that it cancels the off-diagonal part to the first order, so that :
\begin{equation}
\left[S,H_{0}\right]=-H_{v}
\end{equation}
The effective hamiltonian to the second order is given by
\begin{equation}
H_{eff}= H_{0}+\frac{1}{2}\left[S,H_{v}\right]
\end{equation}
The most crucial step in the transformation is to compute the generator of the transformation and insofar the literature reveals no method to get the generator directly from the hamiltonian. In this work, we have come up with an explicit method to calculate the generator of SW transformation. There are other methods to get the effective hamiltonian without recourse to unitary transformation; rather the method is based on projection operators(or Hubbard operators). In this paper we present the SW Transformation of Anderson models in a pedagogical style.

\section{SCHRIEFFER WOLFF TRANSFORMATION OF SIAM}
The generator of SWT for SIAM is given by
\begin{equation}
S=\sum_{k\sigma}(A_{k}+B_{k}n_{d\bar{\sigma}})V_{k}(c^{\dag}_{k\sigma}d_{\sigma}-d^{\dag}_{\sigma}c_{k\sigma}) 
\end{equation}
where   $ A_{k}$ and $ B_{k}$ are given by 
\begin{align}
& A_{k} = \frac{1}{\epsilon_{d}-\epsilon_{k}} \\
& B_{k} = \frac{1}{\epsilon_{d}-\epsilon_{k}+U}-\frac{1}{\epsilon_{d}-\epsilon_{k}} 
\end{align}
To do the SWT we will have to calculate the following commutator:
\begin{equation}
\left[S,H_{v}\right] 
=\left[\sum_{k\sigma}(A_{k}+B_{k}n_{d\bar{\sigma}})V_{k}(c^{\dag}_{k\sigma}d_{\sigma}-d^{\dag}_{\sigma}c_{k\sigma}),\sum_{k'\sigma'}V_{k'}\left(c^{\dag}_{k'\sigma'}d_{\sigma'}+d^{\dag}_{\sigma'}c_{k'\sigma'}\right)\right]  
\end{equation}
Commutators are calculated as described below:
\begin{eqnarray}
&\left[\sum_{k\sigma}A_{k}V_{k}\left(c^{\dag}_{k\sigma}d_{\sigma}-d^{\dag}_{\sigma}c_{k\sigma}\right)
\sum_{k'\sigma'}V_{k'}\left(c^{\dag}_{k'\sigma'}d_{\sigma'}+d^{\dag}_{\sigma'}c_{k'\sigma'}\right)\right] \notag \\
&=\sum_{kk'\sigma}A_{k}V_{k}V_{k'}\left(c^{\dag}_{k\sigma}c_{k'\sigma}+h.c.\right)-\sum_{k\sigma}A_{k}V_{k}^{2}\left(d^{\dag}_{\sigma}d_{\sigma}+h.c.\right) 
\end{eqnarray}

\begin{align}
&\left[\sum_{k\sigma}B_{k}V_{k}n_{d\bar{\sigma}}\left(c^{\dag}_{k\sigma}d_{\sigma}-d^{\dag}_{\sigma}c_{k\sigma}\right),\sum_{k'\sigma'}V_{k'}\left(c^{\dag}_{k'\sigma'}d_{\sigma'}+d^{\dag}_{\sigma'}c_{k'\sigma'}\right)\right] \notag \\ 
&=\sum_{k\sigma}\sum_{k'\sigma'}B_{k}V_{k}V_{k'}\left[n_{d\bar{\sigma}}c^{\dag}_{k\sigma}d_{\sigma},c^{\dag}_{k'\sigma'}d_{\sigma'}+d^{\dag}_{\sigma'}c_{k'\sigma'}\right]-\\
&\left[n_{d\bar{\sigma}}d^{\dag}_{\sigma}c_{k\sigma},c^{\dag}_{k'\sigma'}d_{\sigma'}+d^{\dag}_{\sigma'}c_{k'\sigma'}\right] \notag \\
&=\sum_{k\sigma}\sum_{k'\sigma'}V_{k}V_{k'}\left[n_{d\bar{\sigma}}c^{\dag}_{k\sigma}d_{\sigma},c^{\dag}_{k'\sigma'}d_{\sigma'}\right]+
\left[n_{d\bar{\sigma}}c^{\dag}_{k\sigma}d_{\sigma},d^{\dag}_{\sigma'}c_{k'\sigma'}\right]- \notag\\
&\left[n_{d\bar{\sigma}}d^{\dag}_{\sigma}c_{k\sigma},c^{\dag}_{k'\sigma'}d_{\sigma'}\right]-
\left[n_{d\bar{\sigma}}d^{\dag}_{\sigma}c_{k\sigma},d^{\dag}_{\sigma'}c_{k'\sigma'}\right]])
\end{align}
So the commutator $[S,H_{v}]$ is given by:
\begin{eqnarray}
[S,H_{v}] =&\sum_{kk'\sigma}&A_{k}V_{k}V_{k'}(c^{\dag}_{k\sigma}c_{k'\sigma}+h.c)-\sum_{k\sigma}A_{k}V_{k}^{2}(d^{\dag}_{\sigma}d_{\sigma}+h.c.) - \notag\\
&\sum_{k\sigma}&B_{k}V_{k}^{2}(n_{d\bar{\sigma}}d^{\dag}_{\sigma}+h.c.)+\sum_{kk'\sigma}B_{k}V_{k}V_{k'}(d^{\dag}_{\bar{\sigma}}c_{k'\bar{\sigma}}c^{\dag}_{k\sigma}d_{\sigma}+h.c)-\notag\\
&\sum_{kk'\sigma}&B_{k}V_{k}V_{k'}(c^{\dag}_{k'\bar{\sigma}}d_{\bar{\sigma}}c^{\dag}_{k\sigma}d_{\sigma}+h.c.) + \sum_{kk'\sigma}B_{k}V_{k}V_{k'}(c^{\dag}_{k\sigma}c_{k'\sigma}n_{d\bar{\sigma}}+h.c.) 
\end{eqnarray}
To show that we have got the Kondo exchange term in the above commutator, we will write exchange term in terms of creation and annihilation operators.
\begin{align}
\Psi^{\dag}_{k}&= \begin{pmatrix}
 c^{\dag}_{k\uparrow} \\
c^{\dag}_{k\downarrow}
\end{pmatrix}
& 
\Psi_{k}& = \begin{pmatrix}
c_{k\uparrow} \\
c_{k\downarrow}
\end{pmatrix}
&
\Psi^{\dag}_{d}&=\begin{pmatrix}
d^{\dag}_{\uparrow} \\
d^{\dag}_{\downarrow}
\end{pmatrix}
& 
\Psi_{d}&=\begin{pmatrix}
d_{\uparrow} \\
d_{\downarrow}
\end{pmatrix}
\end{align}
\begin{equation*}
4\left(\Psi^{\dag}_{k}S\Psi_{k'}\right)\left(\Psi^{\dag}_{d}S\Psi_{d}\right)= \Psi^{\dag}_{k}\Psi_{k'}(\sigma.\sigma)\Psi^{\dag}_{d}\Psi_{d}
\end{equation*} 
\hspace{-0.9cm}
\begin{equation*}
= \Psi^{\dag}_{k}\Psi_{k'}(\sigma.\sigma)\Psi^{\dag}_{d}\Psi_{d}
\end{equation*}
\hspace{-0.9cm}
\begin{equation}
= (\Psi^{\dag}_{k}\sigma_{z}\Psi_{k})(\Psi^{\dag}_{d}\sigma_{z}\Psi_{d})+\frac{1}{2}\left((\Psi^{\dag}_{k}\sigma^{+}\Psi_{k})(\Psi^{\dag}_{d}\sigma^{-}\Psi_{d})+(\Psi^{\dag}_{k}\sigma^{-}\Psi_{k})(\Psi^{\dag}_{d}\sigma^{+}\Psi_{d})\right)
\end{equation}


Now calculating the individual terms, we get :
\begin{align}
&(\Psi^{\dag}_{k}\sigma_{z}\Psi_{k'})(\Psi^{\dag}\sigma_{z}\Psi_{d})= \notag\\
& \left[ \begin{pmatrix} c^{\dag}_{k\uparrow} & c^{\dag}_{k\downarrow} \end{pmatrix} \begin{pmatrix} 1 & 0 \\ 0 & -1 \end{pmatrix} \begin{pmatrix} c_{k'\uparrow} \\ c_{k'\downarrow}\end{pmatrix} \right]
\left[ \begin{pmatrix} d^{\dag}_{\uparrow}& d^{\dag}_{\downarrow} \end{pmatrix} \begin{pmatrix} 1 & 0 \\ 0 & -1 \end{pmatrix} \begin{pmatrix} d_{\uparrow} \\ d_{\downarrow} \end{pmatrix}\right]\notag \\
&\left[ \begin{pmatrix} c^{\dag}_{k\uparrow} & c^{\dag}_{k\downarrow} \end{pmatrix} \begin{pmatrix} c_{k'\uparrow} \\ -c_{k'\downarrow}\end{pmatrix} \right]
\left[ \begin{pmatrix} d^{\dag}_{\uparrow} & d^{\dag}_{\downarrow} \end{pmatrix} \begin{pmatrix} d_{\uparrow} \\  - d_{\downarrow} \end{pmatrix}\right]\notag\\
&\left[\begin{pmatrix} c^{\dag}_{k\uparrow} &  c^{\dag}_{k\downarrow}\end{pmatrix} \begin{pmatrix} c_{k'\uparrow} \\ -c_{k'\downarrow} \end{pmatrix} \right]
\left[ \begin{pmatrix} d^{\dag}_{\uparrow} & d^{\dag}_{\downarrow} \end{pmatrix} \begin{pmatrix} d_{\uparrow} \\ -d_{\downarrow} \end{pmatrix} \right]\notag \\
&= \left(c^{\dag}_{k\uparrow}c_{k'\uparrow}-c^{\dag}_{k\downarrow}c_{k'\downarrow}\right)
\left(d^{\dag}_{\uparrow}d_{\uparrow}-d^{\downarrow}d_{\downarrow}\right) \notag\\
&= \sum_{\sigma} \left(c^{\dag}_{k\sigma}c_{k'\sigma}\right)\left(n_{d\sigma}-n_{d\bar{\sigma}}\right) 
\end{align}
\begin{align}
\boxed{\left(\Psi^{\dag}_{k}\sigma_{z}\Psi_{k'}\right)\left(\Psi^{\dag}_{d}\sigma_{z}\Psi_{d}\right)=\sum_{\sigma}c^{\dag}_{k\sigma}c_{k'\sigma}\left(n_{d\sigma}-n_{d\bar{\sigma}}\right)}
\end{align}
\begin{align}
&\left(\Psi^{\dag}_{k}\sigma^{+}\Psi_{k'}\right)\left(\Psi^{\dag}_{d}\sigma^{-}\Psi_{d}\right)=\notag\\
&\left[\begin{pmatrix} c^{\dag}_{k\uparrow} & c^{\dag}_{k\downarrow} \end{pmatrix} \begin{pmatrix} 0 & 1 \\ 0 & 0 \end{pmatrix} \begin{pmatrix} c_{k'\uparrow} \\ c_{k'\downarrow} \end{pmatrix}  \right]
\left[ \begin{pmatrix} d^{\dag}_{\uparrow} & d^{\dag}_{\downarrow} \end{pmatrix} \begin{pmatrix} 0 & 0 \\ 1 & 0 \end{pmatrix} \begin{pmatrix} d_{\uparrow} \\ d_{\downarrow} \end{pmatrix} \right]\notag\\
&\left[ \begin{pmatrix} c^{\dag}_{k\uparrow} & c^{\dag}_{k\downarrow} \end{pmatrix} \begin{pmatrix} c_{k'\downarrow} \\ 0 \end{pmatrix} \right]
\left[ \begin{pmatrix} d^{\dag}_{\uparrow} & d^{\dag}_{\downarrow} \end{pmatrix} \begin{pmatrix} 0 \\ d_{\uparrow} \end{pmatrix} \right]\notag \\
&=c^{\dag}_{k\uparrow}c_{k'\downarrow}d^{\dag}_{\downarrow}d_{\uparrow} 
\end{align}
\begin{align}
\boxed{\left(\Psi^{\dag}_{k}\sigma^{+}\Psi_{k'}\right) \left(\Psi^{\dag}_{d}\sigma^{-}\Psi_{d}\right)=c^{\dag}_{k\uparrow}c_{k'\downarrow}d^{\dag}_{\downarrow}d_{\uparrow}}
\end{align}
Similarly one can calculate
\begin{align}
\boxed{\left(\Psi^{\dag}_{k}\sigma^{+}\Psi_{k'}\right)\left(\Psi^{\dag}_{d}\sigma^{-}\Psi_{d}\right)=c^{\dag}_{k\downarrow}c_{k'\uparrow}d^{\dag}_{\uparrow}d_{\downarrow}}
\end{align}
\begin{align}
\left(\Psi^{\dag}_{k}\sigma\Psi_{k'}\right)\left(\Psi^{\dag}_{d}\sigma\Psi_{d}\right)=\sum_{\sigma}c^{\dag}_{k\sigma}c_{k'\sigma}\left(n_{d\sigma}-n_{d\bar{\sigma}}\right)+\frac{1}{2}\left(c^{\dag}_{k\uparrow}c_{k\downarrow}d^{\dag}_{\downarrow}d_{\uparrow}+h.c\right)
\end{align}
\begin{align}
&4\left(\Psi^{\dag}_{k}S\Psi_{k'}\right)\left(\Psi^{\dag}_{d}S\Psi_{d}\right)=\sum_{\sigma}\left(c^{\dag}_{k\sigma}c_{k'\sigma}\left(n_{d\sigma}-n_{d\bar{\sigma}}\right)+2c^{\dag}_{k\sigma}c_{k'\bar{\sigma}}d^{\dag}_{\bar{\sigma}}d_{\sigma}\right)
\end{align}
Note that in the commutator above, we got the term :
\begin{align}
&\sum_{kk'\sigma}B_{k}V_{k}V_{k'}\left(c^{\dag}_{k\sigma}c_{k'\sigma}n_{d\bar{\sigma}}-c^{\dag}_{k\sigma}c_{k'\bar{\sigma}}d^{\dag}_{\bar{\sigma}}d_{\sigma}\right)+h.c 
\end{align}
\vspace{-0.2cm}
We shall show that this term gives the Kondo exchange term
\begin{align}
&=\sum_{kk'\sigma}B_{k}V_{k}V_{k'}\left(c^{\dag}_{k\sigma}c_{k'\sigma}(\frac{n_{d\sigma}+n_{d\bar{\sigma}}}{2})+c^{\dag}_{k\sigma}c_{k'\sigma}(\frac{n_{d\bar{\sigma}}-n_{d\sigma}}{2})-c^{\dag}_{k\sigma}c_{k'\bar{\sigma}}d^{\dag}_{\bar{\sigma}}d_{\sigma}\right)+h.c. \notag \\
&=\sum_{kk'\sigma}B_{k}V_{k}V_{k'}\left(\frac{-1}{2}\left(c^{\dag}_{k\sigma}c_{k'\sigma}(n_{d\sigma}-n_{d\bar{\sigma}})+c^{\dag}_{k\sigma}c_{k'\bar{\sigma}}d^{\dag}_{\bar{\sigma}}d_{\sigma} \right)+c^{\dag}_{k\sigma}c_{k'\sigma}(\frac{n_{d\sigma}+n_{d_{\bar{\sigma}}}}{2})\right)+h.c.
\end{align}
\vspace{-0.2cm}
So the first two  terms give the exchange term with an additional term generated as well:
\begin{align}
&=- \sum_{kk'} J_{kk'} \left(\Psi^{\dag}_{k}S\Psi_{k'}\right)\left(\Psi^{\dag}_{d}S\Psi_{d}\right)+\sum_{kk'\sigma}B_{k}V_{k}V_{k'}\left( c^{\dag}_{k\sigma}c_{k'\sigma}(\frac{n_{d\sigma}+n_{d\bar{\sigma}}}{2})\right)+h.c
\end{align}
The other terms that also get generated are given below:
\begin{align}
&H_{dir}=\sum_{kk'\sigma}\left(A_{k}V_{k}V_{k'}+B_{k}V_{k}V_{k'}\frac{n_{d\sigma}+n_{d\bar{\sigma}}}{2}\right)c^{\dag}_{k\sigma}c_{k'\sigma}+h.c. \\
&H_{hop}= -\sum_{k\sigma}V_{k}^{2}\left(A_{k}+B_{k}n_{d\bar{\sigma}}\right)n_{d\sigma}+h.c. \\
&H_{ch}=\sum_{kk'\sigma}B_{k}V_{k}V_{k'}\left(c^{\dag}_{k\bar{\sigma}}d_{\bar{\sigma}}c^{\dag}_{k'\sigma}d_{\sigma}\right)+h.c.
\end{align}
\section{SCHRIEFFER WOLFF TRANSFORMATION OF PAM}
The generator of SW transformation for PAM is given by:
\begin{equation}
S= \sum_{k\sigma i}(A_{k}+B_{k}n_{i\sigma})V_{k}(c^{\dag}_{k\sigma}f_{i\sigma}e^{-ikR_{i}}-f^{\dag}_{i\sigma}c_{k\sigma}e^{ikR_{i}})
\end{equation}
where $A_{k}$ and $B_{k}$ are given by 
\begin{align}
A_{k} & = \frac{1}{\epsilon_{k}-\epsilon_{f}}\\
B_{k} & = \frac{1}{\epsilon_{k}-\epsilon_{f}-U}-\frac{1}{\epsilon_{k}-\epsilon_{f}}
\end{align}
To carry out the SW transformation we have to calculate following commutator $[S,H_{v}]=$
\begin{align}
&\left[ \sum_{k\sigma i}(A_{k}+B_{k}n_{i\bar{\sigma}})V_{k}(c^{\dag}_{k\sigma}f_{i\sigma}e^{-ikR_{i}}-f^{\dag}_{i\sigma}c_{k\sigma}e^{ikR_{i}}),
\sum_{k'j\sigma'}V_{k'}(c^{\dag}_{k'\sigma'}f_{j\sigma'}e^{-ik'R_{j}}+f^{\dag}_{j\sigma'}c_{k'\sigma'}e^{ik'R_{j}}\right]
\end{align}
The commutators are calculated in the following way:
\begin{align}
&\left[\sum_{ki\sigma}A_{k}V_{k}c^{\dag}_{k\sigma}f_{i\sigma}e^{ikR_{i}}, 
\sum_{k'j\sigma'}V_{k'}c^{\dag}_{k'\sigma'}f_{j\sigma'}e^{-ik'R_{j}}\right]\notag\\
=&\sum_{ki\sigma}\sum_{k'j\sigma'}A_{k}V_{k}V_{k'}e^{-ikR_{i}}e^{-ik'R_{j}}  \left[c^{\dag}_{k\sigma}f_{i\sigma},c^{\dag}_{k'\sigma'}f_{j\sigma'}\right] \\
=& 0
\end{align}
Similarly  
\begin{align}
\left[ \sum_{ki\sigma}A_{k}V_{k}e^{ikR_{i}}f^{\dag}_{i\sigma}c_{k\sigma},\sum_{jk'\sigma'}V_{k'}e^{-ik'R_{j}}f^{\dag}_{j\sigma'}c_{k'\sigma'}\right]=0
\end{align}

\begin{align}
&\left[\sum_{ki\sigma}A_{k}V_{k}e^{-ikR_{i}}c^{\dag}_{k\sigma}f_{i\sigma},\sum_{k'j\sigma'}V_{k'}e^{ik'R_{j}}f^{\dag}_{j\sigma'}c_{k'\sigma'}\right]\notag\\ 
&= \sum_{ki\sigma}\sum_{k'j\sigma'}A_{k}V_{k'}V_{k}e^{-ikR_{i}} e^{ik'R_{j}}\left[c^{\dag}_{k\sigma}f_{i\sigma},f^{\dag}_{j\sigma'}c_{k'\sigma'}\right] \notag\\
&=\sum_{ki\sigma}\sum_{k'\sigma'}A_{k}V_{k'}V_{k} e^{i(k'-k)R_{i}}c^{\dag}_{k\sigma}c_{k'\sigma}- 
 \sum_{ijk\sigma}A_{k}V^{2}_{k}e^{ik(R_{j}-R_{i})}f^{\dag}_{j\sigma}f_{i\sigma} 
\end{align}
Hermitian conjugate of above commutator is given by:
\begin{align}
&\left[\sum_{ik\sigma}A_{k}V_{k}f^{\dag}_{i\sigma}c_{k\sigma}e^{ikR_{i}},\sum_{jk'\sigma'}c^{\dag}_{k'\sigma'}f_{j\sigma'}e^{-ik'R_{j}}\right]\notag\\
=& \sum_{ki\sigma}\sum_{jk'\sigma'}A_{k}V_{k}V_{k'}e^{ikR_{i}}e^{-ik'R_{j}} \left[f^{\dag}_{i\sigma}c_{k\sigma},c^{\dag}_{k'\sigma'}f_{j\sigma'}\right]\notag\\
=&\sum_{ijk\sigma}A_{k}V^{2}_{k} e^{ik(R_{i}-R_{j})}f^{\dag}_{i\sigma}f_{j\sigma}+ 
 \sum_{kk'i\sigma}A_{k}V_{k}V_{k'}e^{i(k-k')R_{i}}c^{\dag}_{k'\sigma}c_{k\sigma}
\end{align}
\vspace{-0.5cm}
\begin{align}
&\left[\sum_{ki\sigma} B_{k}V_{k} n_{i\bar{\sigma}} c^{\dag}_{k\sigma}f_{i\sigma}e^{-ikR_{i}},
\sum_{k'j\sigma'}V_{k'}c^{\dag}_{k'\sigma'}f_{j\sigma'}e^{-ik'R_{j}}\right] \notag\\
& = \sum_{ki\sigma}\sum_{jk'\sigma'}V_{k}V_{k'}e^{-ikR_{i}}e^{-ik'R_{j}} \left[n_{i\bar{\sigma}} c^{\dag}_{k\sigma}f_{i\sigma},c^{\dag}_{k'\sigma'}f_{j\sigma'}\right]\notag\\
&=\sum_{kk'i\sigma}B_{k}V_{k}V_{k'}e^{-i(k+k')R_{i}}c^{\dag}_{k'\bar{\sigma}}c^{\dag}_{k\sigma}f_{i\bar{\sigma}}f_{i\sigma}
\end{align}
\begin{align}
&\left[\sum_{ki\sigma}B_{k}V_{k}n_{i\bar{\sigma}}f^{\dag}_{i\sigma}c_{k\sigma}e^{ikR_{i}},
\sum_{jk'\sigma'}V_{k'}f^{\dag}_{j\sigma'}c_{k'\sigma'}e^{ik'R_{j}}\right] \notag\\
&=\sum_{ik\sigma}\sum_{jk'\sigma'}B_{k}V_{k}V_{k'}e^{ikR_{i}}e^{ik'R_{j}}\left[n_{i\bar{\sigma}}f^{\dag}_{i\sigma}c_{k\sigma},
f^{\dag}_{j\sigma'}c_{k'\sigma'}\right] \notag\\
&= \sum_{k'ki\sigma}B_{k}V_{k}V_{k'}e^{i(k+k')R_{i}}f^{\dag}_{i\bar{\sigma}}f^{\dag}_{i\sigma}c_{k'\bar{\sigma}}c_{k\sigma}
\end{align}
\hspace{-1cm}

\begin{align}
&\left [ \sum_{ik\sigma} B_{k}V_{k}n_{i\bar{\sigma}}c^{\dag}_{k\sigma}f_{i\sigma}e^{-ikR_{i}},\sum_{jk'\sigma'}V_{k'}f^{\dag}_{j\sigma'}c_{k'\sigma'}e^{ik'R_{j}}\right]\notag \\
&=\sum_{ik\sigma}\sum_{jk'\sigma'}B_{k}V_{k}V_{k'}e^{-ikR_{i}}e^{ik'R_{j}}\left[n_{i\bar{\sigma}}c^{\dag}_{k\sigma}f_{i\sigma},f^{\dag}_{j\sigma'}c_{k'\sigma'}\right]\notag\\ &=\sum_{ikk'\sigma}B_{k}V_{k}V_{k'}e^{i(k'-k)R_{i}}n_{i\bar{\sigma}}c^{\dag}_{k\sigma}c_{k'\sigma}- \sum_{ijk\sigma}B_{k}V_{k}^{2}e^{ik(R_{j}-R_{i})}n_{i\bar{\sigma}}f^{\dag}_{j\sigma}f_{i\sigma} + \notag\\
&\hspace{0.6cm}\sum_{ikk'\sigma}B_{k}V_{k}V_{k'}e^{i(k'-k)R_{i}}f^{\dag}_{i\bar{\sigma}}c_{k\bar{\sigma}}c^{\dag}_{k\sigma}f_{i\sigma} 
\end{align}

\begin{align}
&\left[ \sum_{ik\sigma} B_{k}V_{k}n_{i\bar{\sigma}}f^{\dag}_{i\sigma}c_{k\sigma}e^{ikR_{i}},\sum_{jk'\sigma'}V_{k'}c^{\dag}_{k'\sigma'}f_{j\sigma'}e^{-ik'R_{j}} \right] \notag\\
&=\sum_{ik\sigma}\sum_{jk'\sigma'}B_{k}V_{k}V_{k'}e^{ikR_{i}}e^{-ik'R_{j}} \left[n_{i\bar{\sigma}}f^{\dag}_{i\sigma}c_{k\sigma},c^{\dag}_{k'\sigma'}f_{j\sigma'}\right] \notag\\
&=\sum_{ikk'\sigma}B_{k}V_{k}V_{k'}e^{i(k-k')R_{i}}n_{i\bar{\sigma}}c^{\dag}_{k'\sigma}c_{k\sigma} 
-\sum_{ijk\sigma}B_{k}V_{k}^{2}e^{ik(R_{i}-R_{j})}n_{i\bar{\sigma}}f^{\dag}_{i\sigma}f_{j\sigma}+ \notag\\
&\hspace{0.6cm}\sum_{ikk'\sigma}e^{i(k-k')R_{i}}c^{\dag}_{k'\bar{\sigma}}f_{i\bar{\sigma}}f^{\dag}_{i\sigma}c_{k\sigma} 
\end{align}

Combining all these commutators, we get the final commutator as :
\vspace{0.3cm}
\begin{align}
&\left[ \sum_{k\sigma i}(A_{k}+B_{k}n_{i\bar{\sigma}})V_{k}(c^{\dag}_{k\sigma}f_{i\sigma}e^{-ikR_{i}}-f^{\dag}_{i\sigma}c_{k\sigma}e^{ikR_{i}}),
\sum_{k'j\sigma'}V_{k'}(c^{\dag}_{k'\sigma'}f_{j\sigma'}e^{-ik'R_{j}}+f^{\dag}_{j\sigma'}c_{k'\sigma'}e^{ik'R_{j}}\right]\notag\\
&=\sum_{ki\sigma}\sum_{k'j'\sigma'}(A_{k}V_{k}V_{k'}e^{i(k'-k)R_{i}}c^{\dag}_{k\sigma}c_{k'\sigma}+h.c.)-\sum_{ijk\sigma}(A_{k}V_{k}^{2}e^{ik(R_{j}-R_{i})}f^{\dag}_{j\sigma}f_{i\sigma}+h.c.)+\notag\\
&\sum_{kk'i\sigma}(B_{k}V_{k}V_{k'}e^{i(k+k')R_{i}}c^{\dag}_{k'\bar{\sigma}}c_{k\sigma}f_{i\bar{\sigma}}f_{i\sigma}+h.c.)+ \sum_{ikk'\sigma}B_{k}V_{k}V_{k'}e^{i(k'-k)R_{i}}n_{i\bar{\sigma}}c^{\dag}_{k\sigma}c_{k'\sigma} - \notag\\
&\sum_{ijk\sigma}(B_{k}V_{k}^{2}e^{ik(R_{j}-R_{i})}n_{i\bar{\sigma}}f^{\dag}_{j\sigma}f_{i\sigma}+h.c.)+\sum_{ikk'\sigma}(B_{k}V_{k}V_{k'}e^{i(k'-k)R_{i}}f^{\dag}_{i\sigma}c_{k'\bar{\sigma}}c^{\dag}_{k\sigma}f_{i\sigma}+h.c.)
\end{align}
\newpage
We see that there are many terms that comes out of SW transformation. One important term which gets produced, is Kondo exchange term. Proceeding in a similar way as was done for SIAM, we obtain the Kondo exchange term from the following terms of the commutator $ [S,H_{v}]$.
\begin{align}
&\sum_{kk'i\sigma}B_{k}V_{k}V_{k'}e^{i(k-k')R_{i}}\left(n_{i\bar{\sigma}}c^{\dag}_{k'\sigma}+c^{\dag}_{k'\bar{\sigma}}f_{i\bar{\sigma}}f^{\dag}_{i\sigma}c_{k\sigma}\right) \notag \\
&=\sum_{kk'i\sigma}B_{k}V_{k}V_{k'}e^{i(k-k')R_{i}}(\frac{1}{2}\left(n_{i\sigma}+n_{i\bar{\sigma}}\right)c^{\dag}_{k'\sigma}c_{k\sigma}-
\frac{1}{2}\left(n_{i\sigma}-n_{i\bar{\sigma}}\right)c^{\dag}_{k'\sigma}c_{k\sigma}+ c^{\dag}_{k'\bar{\sigma}}f_{i\bar{\sigma}}f^{\dag}_{i\sigma}c_{k\sigma}) \notag\\
&=\sum_{kk'i\sigma}B_{k}V_{k}V_{k'}e^{i(k-k')R_{i}} \left(\frac{1}{2}\left(n_{i\sigma}+n_{i\bar{\sigma}}\right)c^{\dag}_{k'\sigma}c_{k\sigma}\right) - 
2\sum_{kk'i}B_{k}V_{k}V_{k'}\left(\Psi^{\dag}_{k'}S\Psi_{k}\right)\left(\Psi^{\dag}_{f}S\Psi_{f}\right) 
\end{align}
 The exchange term comes from the second term of the above equation
\begin{equation}
H_{ex}=\sum_{kk'i}J_{k'k}e^{-i(k'-k)R_{i}}\left(\Psi^{\dag}_{k'}S\Psi_{k}\right)\left(\Psi^{\dag}_{f}S\Psi_{f}\right) 
\end{equation}
where $J_{k'k}$ is given by
\begin{equation}
 J_{k'k}=V_{k}V_{k'}(\frac{-1}{\epsilon_{k}-\epsilon_{f}-U}-\frac{1}{\epsilon_{k'}-\epsilon_{f}-U}+\frac{1}{\epsilon_{k}-\epsilon_{f}}+\frac{1}{\epsilon_{k'}-\epsilon_{f}})
\end{equation}
\begin{align}
&H_{dir}=\sum_{kk'i\sigma}A_{k}V_{k}V_{k'}e^{-i(k'-k)R_{i}}c^{\dag}_{k'\sigma}c_{k\sigma}+\left(B_{k}V_{k}V_{k'}\frac{1}{2}\left(n_{i\sigma}+n_{i\bar{\sigma}}\right)c^{\dag}_{k'\sigma}c_{k\sigma}\right) \\
&H_{dir}=\sum_{kk'i\sigma}\left(W_{kk'}-\frac{1}{4}J_{kk'}\left(n_{i\sigma}+n_{i\bar{\sigma}}\right)e^{-i(k'-k)R_{i}}c^{\dag}_{k'\sigma}c_{k\sigma}\right) 
\end{align}
where $ W_{kk'}$ is given by
\begin{equation}
W_{kk'} = \frac{1}{2}V_{k'}V_{k}\left(\frac{1}{\epsilon_{k}-\epsilon_{f}}+\frac{1}{\epsilon_{k'}-\epsilon_{f}}\right) 
\end{equation}
\begin{align}
H_{hop}=-&\sum_{ijk\sigma}B_{k}V^{2}_{k}(e^{ik(R_{i}-R_{j})}n_{i\bar{\sigma}}f^{\dag}_{i\sigma}f_{j\sigma} - 
e^{ik(R_{j}-R_{i})}n_{i\bar{\sigma}}f^{\dag}_{j\sigma}f_{i\sigma}) -\notag \\
&\sum_{ijk\sigma}A_{k}V_{k}^{2}e^{ik(R_{j}-R_{i})}f^{\dag}_{j\sigma}f_{i\sigma} \notag\\
=-&\sum_{ijk\sigma}W_{kk'}-\frac{1}{4}J_{kk'}\left(n_{i\bar{\sigma}}+n_{j\bar{\sigma}}\right)e^{-ik(R_{i}-R_{j})}f^{\dag}_{j\sigma}f_{i\sigma} 
\end{align}
\begin{align}
&H_{ch}=\sum_{ikk'\sigma}B_{k}V_{k}V_{k'}e^{-i(k+k')R_{i}}c^{\dag}_{k'\bar{\sigma}}c^{\dag}_{k\sigma}f_{i\bar{\sigma}}f_{i\sigma} +h.c.
\end{align}
\section{CONCLUSION}
Schrieffer-Wolff transformation is a very important transformation both in Quantum Mechanics and Quantum many-body physics.It is used to get the effective hamiltonian of a given hamiltonian by integrating out high energy degrees of freedom. It is also used for the diagonalization of the hamiltonians. In quantum mechanics, it is the method to perform the degenerate perturbation theory calculations. In this paper, we have presented the detailed calculations of Schrieffer-Wolff transformation of Anderson Models in a pedagogical manner. SW transformation can be carried out for other models as well, following the method given in this paper.
\begin{acknowledgements}
The authors acknowledge DST for financial support and JNCASR for conducive research environment.
\end{acknowledgements}

\end{document}